\documentclass[twocolumn,english,prl,amsmath,amssymb,preprintnumbers]{revtex4}
\usepackage[T1]{fontenc}
\usepackage[latin9]{inputenc}
\usepackage{amstext}
\usepackage{amsmath}
\usepackage{amsfonts}
\usepackage{amssymb}
\usepackage{mathtools}
\usepackage{graphicx}
\usepackage{rotating}
\usepackage {bm}
\usepackage[section]{placeins}
\usepackage{color}
\usepackage{multirow}
\usepackage{cellspace}
\makeatletter
\@ifundefined{textcolor}{}
{%
 \definecolor{BLACK}{gray}{0}
 \definecolor{WHITE}{gray}{1}
 \definecolor{RED}{rgb}{1,0,0}
 \definecolor{GREEN}{rgb}{0,1,0}
 \definecolor{BLUE}{rgb}{0,0,1}
 \definecolor{CYAN}{cmyk}{1,0,0,0}
 \definecolor{MAGENTA}{cmyk}{0,1,0,0}
 \definecolor{YELLOW}{cmyk}{0,0,1,0}
}

\def\kF{k_{\text{F}}}
\def\vF{v_{\text{F}}}
\def\TF{T_{\text{F}}}
\def\kB{k_{\text{B}}}
\def\muB{\mu_{\text{B}}}
\def\NF{N_{\text{F}}}

\def\TC{$T_{\text{C}}$}

\def\ne{n_{\text{e}}}
\def\vso{v_{\text{so}}}
\def\Eso{E_{\text{so}}}

\def\sgn{{\text{sgn\,}}}

\def\be{\begin{equation}}
\def\ee{\end{equation}}
\def\bea{\begin{eqnarray}}
\def\eea{\end{eqnarray}}
\def\bse{\begin{subequations}}
\def\ese{\end{subequations}}

\usepackage{dcolumn}
\usepackage{bm}
\usepackage{amsfonts}
\draft

\makeatother

\usepackage{babel}
\begin{document}
\preprint{Phys. Rev. Lett. {\bf 124}, 147201 (2020)}

\title{Ferromagnetic Quantum Critical Point in Non-Centrosymmetric Systems}

\author{T. R. Kirkpatrick$^{1}$ and  D. Belitz$^{2,3}$}

\affiliation{$^{1}$ Institute for Physical Science and Technology, University of Maryland, College Park, MD 20742, USA\\
                 $^{2}$ Department of Physics and Institute of Theoretical Science, University of Oregon, Eugene, OR 97403, USA\\
                 $^{3}$ Materials Science Institute, University of Oregon, Eugene, OR 97403, USA
                  }

\date{\today}
\begin{abstract}
Ferromagnetic quantum criticality in clean metals has proven elusive due to fermionic soft modes that drive the
transition first order. We show that non-centrosymmetric metals with a strong spin-orbit interaction provide
a promising class of materials for realizing a ferromagnetic quantum critical point in clean systems. The spin-orbit interaction renders
massive the soft modes that interfere with quantum criticality in most materials, while the absence of spatial inversion 
symmetry precludes the existence of new classes of soft modes that could have the same effect. 
\end{abstract}
%
%
\maketitle
Ferromagnetism in metals has provided one of the earliest examples of a quantum phase transition. 
Stoner \cite{Stoner_1938} developed the eponymous mean-field theory that describes both the classical 
and the quantum ferromagnetic (FM) transition. Hertz \cite{Hertz_1976} later argued that, in the quantum 
case (i.e., for the transition at zero temperature driven by a non-thermal control parameter) Stoner theory 
is exact, as far as the critical behavior is concerned, for all spatial dimensions $d>1$.
The reason is that the coupling between the statics and the dynamics at zero temperature ($T=0$) lowers
the upper critical dimension, above which the fluctuations neglected by mean-field theory are irrelevant,
from $d_c^+ = 4$ in the classical case to $d_c^+ = 1$ in the quantum case. Hertz's renormalization-group 
(RG) treatment, as refined by Millis \cite{Millis_1993}, agreed with results obtained by different methods by
Moriya \cite{Moriya_1985}. Collectively, this became known as the Hertz-Millis-Moriya (HMM) theory of
FM quantum criticality.

These theoretical predictions were not borne out experimentally. In stoichiometric systems with minimal
amounts of quenched disorder, and the quantum phase transition (QPT) driven by pressure, the 
transition almost invariably becomes first order if the Curie temperature is sufficiently low,
and this is true for local-moment ferromagnets as well as for itinerant ones \cite{Brando_et_al_2016a}.
Notable exceptions are CeRh$_6$Ge$_4$ \cite{Kotegawa_et_al_2019, Shen_et_al_2019} and UIr \cite{Kobayashi_et_al_2006},
which we will come back to. The reason for this failure of HMM theory is by know well known: A generic
Fermi liquid, with a negligible spin-orbit interaction (more on this later), contains soft or massless two-particle
excitations that couple to, and are rendered massive by, an external magnetic field or a 
magnetization \cite{Belitz_Kirkpatrick_Vojta_1997, Betouras_Efremov_Chubukov_2005}.
This coupling results in the free energy being a nonanalytic function of the magnetization, which
in turn drives the FM QPT first order \cite{Belitz_Kirkpatrick_Vojta_1999}. This mechanism
is operative for local-moment ferromagnets as well as for itinerant ones, and also for canted ferromagnets
and for ferrimagnets \cite{Kirkpatrick_Belitz_2012b} as well as for magnetic nematics \cite{Kirkpatrick_Belitz_2011};
for a review, see Ref.~\onlinecite{Brando_et_al_2016a}.
It has recently been shown that it also is operative in  Dirac metals, i.e., systems where a linear band crossing
is caused by a strong spin-orbit coupling \cite{Kirkpatrick_Belitz_2019a, Kirkpatrick_Belitz_2019b, Belitz_Kirkpatrick_2019}, if
for unobvious reasons. A nonzero temperature gives the soft modes a mass and thus cuts off the first-order mechanism; this
leads to a tricritical point in the phase diagram \cite{Belitz_Kirkpatrick_Vojta_1999}. Similarly, an
external magnetic field gives the soft modes a mass, which results in tricritical wings that emerge
from the tricritical point in the temperature-pressure-field parameter space and end in quantum
critical points (QCPs) at a nonzero field \cite{Belitz_Kirkpatrick_Rollbuehler_2005}. 

One way to avoid these conclusions, and realize a FM QCP in zero field,
is to introduce  quenched disorder, which has
been predicted \cite{Belitz_Kirkpatrick_Vojta_1999, Sang_Belitz_Kirkpatrick_2014} and observed
\cite{Goko_et_al_2017} to restore a QCP. However, the resulting critical behavior is not described by 
HMM theory, but is substantially more complicated \cite{Kirkpatrick_Belitz_1996, Belitz_et_al_2001b, Kirkpatrick_Belitz_2014}.
Experimental results are consistent with these predictions \cite{Huang_et_al_2016, Sales_et_al_2017}.
Another possibility are one-dimensional or quasi-one-dimensional materials, see Ref.~\onlinecite{Komijani_Coleman_2018}
for a model of FM quantum criticality in such systems.

It would be very interesting if clean materials could be found in which the mechanism
for a first-order transition is inoperative, so that a FM QCP in zero field can
be realized in three-dimensional systems. 
In this Letter we show that a promising class of materials 
are systems with a
strong spin-orbit coupling that are not centrosymmetric. Our central result is an equation of
state that takes the form
\be
h = r\,m - v\,m^3\,\ln\left(\frac{1}{m^2 + \nu^2 + t^2}\right) + u\,m^3\ .
\label{eq:1}
\ee
Here $m$, $\nu$, and $h$ are the dimensionless magnetization, spin-orbit coupling, and magnetic field, respectively,
in atomic units. They are formally defined as follows. Let $\mu$ be the magnetization measured in units of $\muB$ per
volume and $E_{\text{ex}}$ the exchange splitting due to that magnetization,
 $H$ the external magnetic field, $\Eso$ the splitting of the conduction band near the Fermi energy
induced by the spin-orbit coupling, $\ne$ the conduction-electron density, and $\TF$ the Fermi temperature. 
Then $h = \muB H/\kB \TF$,  $m = \mu/\ne \approx E_{\text{ex}}/\kB\TF$, and $\nu = \Eso/\kB\TF$. $t=T/T_0$ is the dimensionless temperature,
with $T_0$ a temperature scale that depends on microscopic details such as the band structure and the correlation 
strength. $r$ is the control parameter, and $u>0$ and $v>0$ are Landau parameters. $u$ is generically of order
unity. $v$ is a measure of the strength of correlations in the system; for very strong correlations, $v\alt 0.1$.

We first discuss Eq.~(\ref{eq:1}) in the context of the general FM QPT problem
and give plausibility arguments for its functional form, then we discuss its implications, and finally we
sketch its derivation.

To make Eq.~(\ref{eq:1}) plausible,  consider the case of a vanishing spin-orbit coupling, $\nu=0$.
Then we recover the equation of state that has been discussed before \cite{Belitz_Kirkpatrick_Vojta_1999,
Brando_et_al_2016a}. The nonanalytic dependence of the free energy, and hence the equation of state,
on the order parameter $m$ at $T=0$ is the result of ballistic soft modes that have been integrated out in order
to express the free energy entirely in terms of the order parameter. The nonanalytic term dominates the
quartic term in the free energy (or the cubic term in the equation of state), and its sign is negative, which
leads to a first-order transition at $r = r_1 = v\,e^{-(1+u/v)}$
where the magnetization changes discontinuously from zero to
$m_1 = e^{-(1+u/v)/2}$ \cite{Belitz_Kirkpatrick_Vojta_1999}.  A nonvanishing temperature gives the soft 
modes a mass, so $T>0$ cuts off the singularity. As a result, there is a tricritical point  at a temperature 
$T_{\text{tc}} = T_0\,e^{-u/2v}$ \cite{Belitz_Kirkpatrick_Vojta_1999}. In a magnetic field, tricritical wings 
emerge from the tricritical point that end in wing tips at $T=0$ and $h = h_{\text{c}} = (4/3)v e^{-3u/2v - 13/4}$ 
\cite{Belitz_Kirkpatrick_Rollbuehler_2005}. 

\begin{figure}[t]
\includegraphics[width=7.5cm]{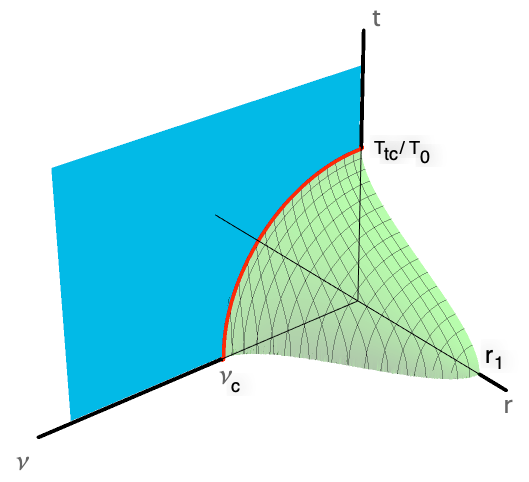}
\caption{Phase diagram in the space spanned by $t$, $r$, and $\nu$ based on Eq.~(\ref{eq:1}). Shown are a flat surface of
              second-order transitions at $r=0$ (solid, blue), a curved surface of first-order transitions (meshed, green), and a line
              of tricritical points delineating the two (red). In a given material $\nu$ will be fixed. For a given $\nu$, $r$ will 
              be a complicated function of pressure and temperature.}
\label{fig:1}
\end{figure}

A spin-orbit interaction splits the conduction band and gives the soft modes a mass. However, in
centrosymmetric systems a chiral degree of freedom leads to new soft modes that have the same
effect as the original ones. Such metals were called Dirac metals in 
Refs.~\onlinecite{Kirkpatrick_Belitz_2019a, Kirkpatrick_Belitz_2019b, Belitz_Kirkpatrick_2019} in order to distinguish them from
the ordinary, or Landau, metals with a negligible spin-orbit interaction.
The net result is an equation of state that is again given by Eq.~(\ref{eq:1})
with $\nu = 0$. This changes if spatial inversion 
symmetry is broken. The spin-orbit interaction still gives the soft modes a mass, but there is no chiral 
degree of freedom that leads to a new class of soft modes. One then obtains
Eq.~(\ref{eq:1}); the resulting phase diagram is shown in Fig.~\ref{fig:1}. 

%

We now give a semi-quantitative discussion of Eq.~(\ref{eq:1}), with the goal of identifying promising
candidate materials that might realize a FM QCP in clean systems. The critical
value $\nu_{\text{c}}$ of the dimensionless spin-orbit energy, above which the first-order transition is 
suppressed, is obviously the same as the dimensionless tricritical temperature $t_{\text{tc}}^0$ for $\nu=0$:
$\nu_{\text{c}} = e^{-u/2v}$. The tricritical temperature $T_{\text{tc}}^0$ for centrosymmetric materials, where $\nu$ is
absent in the equation of state, is typically on the order of 10\,K  \cite{Brando_et_al_2016a}, albeit with
a large spread that ranges from 1\,K in URhGe to over 100\,K in CoS$_2$. For the critical spin-orbit
energy we thus have $\Eso^c = \kB T_0\, \nu_{\text{c}} = (\TF/T_0)\kB T_{\text{tc}}^0$. The temperature scale
$T_0$ has been estimated in Ref.~\onlinecite{Sang_Belitz_Kirkpatrick_2014}, which concluded that
$\TF/T_0$ is typically on the order of 1,000 (as low as 750 in ZrZn$_2$ and as high as 3,500 in UGe$_2$). 
This implies that typically $\Eso^{\text{c}} \approx 1\,$eV, with a spread of a factor of up to 10 in either direction. 
If we assume $\TF \approx 10^5\,$K for a good metal, this implies $\Eso^c/\kB\TF \approx 0.1$. It is illustrative
to compare this with values of $m_1$ for $\nu=0$, which typically fall into a range $m_1 \approx 0.05 - 0.25$
\cite{Brando_et_al_2016a, Sang_Belitz_Kirkpatrick_2014}. 

\begin{table*}[t]
\caption{Non-centrosymmetric pressure-tuned quantum ferromagnets. 
               \TC\ = Curie temperature. $T_{\text{tc}} =$ tricritical temperature.
                  $\Eso$ = spin-orbit splitting of conduction band.
                      $\rho_0 =$ residual resistivity. n.a. = not available. N/A = not applicable. NPT = No pressure tuning to date.
                         AFM =antiferromagnet.
                            References are given only for properties not referenced in Ref.~\onlinecite{Brando_et_al_2016a}.}
\vskip 2pt
\begin{ruledtabular}
\begin{tabular}{lllllllll}
System \ \ \ 
   &Space 
      & Order
         & \TC/K$\,^c$
            & $T_{\text{tc}}$/K
               & $\Eso$/eV
                     & Disorder$\,^d$  
                        & Comments \\
   & Group$^a$
      &of QPT$\,^{b}$ 
         &                            
            & 
               &                              
                     & ($\rho_0/\mu\Omega$cm)\ \ 
                        & \\
\hline\\[-7pt]
MnSi
   & P2$_1$3 (198)
      & 1st
         & 29.5
            & $\approx 10$ 
               &  $\approx 0.3$ \cite{MnSi_footnote}
                     & 0.33
                        & weak helimagnet\\
\\[-7pt]
U$_3$P$_4$
   & I$\bar 4$3d (220)
      & 1st 
         & 138
            & 32
               & n.a.
                     & 4
                        & \\
 \\[-7pt]
 UCoAl
    &P$\bar 6$2m (189)
       & 1st
          & 0
             & >11
                & n.a.
                   & 24
                      &     \\
 \\[-7pt]
 URhAl
    &P$\bar 6$2m (189)
       & 1st
          & $\approx 30$
             & $\approx 11$
                & $\approx 1$ \cite{Kunes_et_al_2001}
                   & $\approx 65$
                      &     \\                    
 \\[-7pt]
SmNiC$_2$
    &Amm2 (38)
       & 1st
          & $\approx 4$
             & >17 (?)
                & n.a.
                   & 2
                      &     \\                    
 \\[-7pt]
\hline\\ [-5pt]
UIr
    &P2$_1$ (4)
       & 2nd
          & 46 - 1
             & N/A
                & n.a.
                   & 0.4 \cite{Akazawa_et_al_2004}
                      & Multiple FM phases. 2nd order above $0.8\,$K    \\                    
 \\[-7pt]
 CeRh$_6$Ge$_4$ 
    &P$\bar 6$m2 (187)
       & 2nd
          & 
             & N/A
                & n.a.
                   & 1.5
                      & QCP with NFL transport behavior \cite{Kotegawa_et_al_2019, Shen_et_al_2019} \\                    
 \\[-7pt]
\hline\\ [-5pt]
Sm$_2$Fe$_{12}$P$_7$
   & P$\bar 6$ (174) 
      & n.a.
         & 6.3~\cite{Janoschek_et_al_2011}
            & N/A
               & n.a.
                  & 6~\cite{Janoschek_et_al_2011}
                     &  NPT \\
\\[-7pt]                     
CePt$_3$B
   & P4mm (99)
      & n.a.
         & 6 \cite{Rauch_et_al_2012}
            & n.a.
               & n.a.
                  & n.a.
                     &NPT.  AFM phase between 7.8K and 6K \\ 
\\[-7pt]
CePdSi$_3$
   & I4mm (107)
      & n.a.
         & 2.78 \cite{Ueta_Ikeda_Yoshizawa_2016}
            & n.a.
               & n.a.
                  & n.a.
                     & NPT. Multiple magnetic phases \\ 
\\[-7pt]
UPtAl
   &P$\bar 6$2m (189)
      & n.a.
         &42.5 \cite{Honda_et_al_2002}
            & N/A
               & $\approx 1$ \cite{Andreev_et_al_2001}
                  & n.a.
                     & \TC\ increases with pressure, no QPT observed \\ 
\\[-7pt]
CeNiC$_2$
   & Amm2 (38)
      & n.a.
         & $\approx 2$ \cite{Katano_et_al_2019}
            & N/A
               & n.a.
                  & $\approx 10$  \cite{Katano_et_al_2019}
                     & Transition to AFM under pressure \\ 
\\[-7pt]
\hline\hline\\[-5pt]
\multicolumn{8}{l} {$^a$ International Short Symbol (Number Index) from https://materials.springer.com}\\
\multicolumn{8}{l} {$^{b}$ At the lowest \TC\ achieved.}\\
\multicolumn{8}{l} {$^{c}$  At ambient pressure for systems with a 1st order QPT;  range as a function of pressure for systems with a 2nd order QPT.}\\      
\multicolumn{8}{l} {$^{d}$  For the highest-quality samples.}\\      [-0pt]
\end{tabular}
\end{ruledtabular}
\vskip -0mm
\label{table:1}
\end{table*}

Considering the two entries in Table~\ref{table:1} with a first-order QPT for which $\Eso$ is known, MnSi and URhAl, it
is plausible that $\Eso$ is not large enough to suppress the first-order mechanism. For the two entries
with a QCP, UIr and CeRh$_6$Ge$_4$,  the spin-orbit splitting $\Eso$ is not known. For a list of $\Eso$ values
in non-centrosymmetric materials that are not ferromagnetic, see Ref.~\onlinecite{Smidman_et_al_2017};
they range from 0.004 eV to 0.2 eV.  For interpreting these values it is important to keep in mind that $\Eso$ should 
be compared to the Fermi energy. For instance, in BiTeBr $\Eso \approx \kB\TF$ \cite{Tokura_Nagaosa_2018}. 
$\Eso \approx 0.2\,$eV has been reported for CePt$_3$Si \cite{Samokhin_Zijlstra_Bose_2004}, which also is not ferromagnetic.
 If the spin-orbit coupling in CeRh$_6$Ge$_4$ were of similar strength, then it would be in the lower range of values that can
plausibly be expected to be responsible for the observed QCP. In UIr one would expect an even
higher value, which may well be the reason for the observed QCP. 

Of the third group of materials listed in Table~\ref{table:1}, the first three are potential candidates
for a pressure-induced QCP, but the values of $\Eso$ are not known. More generally, we conclude that
promising candidates for a FM QCP are non-centrosymmetric materials with a large ($\approx 1\,$eV
or larger) spin-orbit splitting $\Eso$ of the conduction band near the Fermi energy.


We now sketch the derivation of Eq.~(\ref{eq:1}); see the Supplemental Material for more details. In the
absence of spatial inversion symmetry a single-particle Hamiltonian that captures the dominant
effects of the spin-orbit interaction can be written~\cite{Dyakonov_2008}
\be
H_0 = \xi_{\bm k}\,\sigma_0 + \vso\, {\bm\sigma}\cdot{\bm\Omega}({\bm k}) - {\bm h}\cdot{\bm\sigma}\ .
\label{eq:2}
\ee
Here $\xi_{\bm k} = \epsilon_{\bm k} - \mu$ with $\epsilon_{\bm k}$ the single-particle energy-momentum relation 
and $\mu$ the chemical potential, ${\bm h}$ is an external magnetic field, $\bm\sigma = (\sigma_1,\sigma_2,\sigma_3)$ denotes
the Pauli matrices with $\sigma_0$ the $2\times 2$ unit matrix, and $\vso$ is a coupling constant
that represents the strength of the spin-orbit interaction. Invariance under time reversal (which flips the signs
of both $\bm\sigma$ and $\bm k$) in the absence of a magnetic field requires ${\bm\Omega}(-{\bm k}) = -{\bm\Omega}({\bm k})$.
This implies that the spin-orbit term is not invariant under spatial inversion (which flips the sign of $\bm k$ only). 

The explicit form of ${\bm\Omega}({\bm k})$ depends on the space group; well-known examples are 
the Dresselhaus spin-orbit coupling for the zincblende structure, which is cubic in ${\bm k}$ 
\cite{Dresselhaus_1955}, and the Rashba-Sheka coupling for the wurtzite structure, which is linear in 
${\bm k}$~\cite{Rashba_Sheka_1959}. For definiteness, we will use the same form as in
Refs.~\cite{Kirkpatrick_Belitz_2019a, Kirkpatrick_Belitz_2019b, Belitz_Kirkpatrick_2019}, namely, 
\be
\bm\Omega({\bm k}) = {\bm k}\ .
\label{eq:3}
\ee
The coupling constant $\vso$ then is dimensionally a velocity. The broken inversion symmetry is the crucial difference 
between the current discussion and the Dirac metals considered 
in Refs.~\cite{Kirkpatrick_Belitz_2019a, Kirkpatrick_Belitz_2019b, Belitz_Kirkpatrick_2019}. Spatial inversion symmetry requires the
existence of an additional, chiral, pseudo-spin degree of freedom that is odd under parity. The presence or
absence of this degree of freedom qualitatively changes the soft-mode spectrum of the electron system,
as we will now discuss.

The inverse Green function for the Hamiltonian $H_0$ in Eq.~(\ref{eq:2}) is $G_k^{-1} = i\omega_m\, \sigma_0 - H_0$,
with $\omega_m$ a fermionic Matsubara frequency and $k=(i\omega_n,{\bm k})$. In terms of quasiparticle resonances
\bse
\label{eqs:4}
\be
F_k^{\beta} = 1/\left(i\omega_n - \xi_{\bm k} - \beta \vert v{\bm k}-{\bm h}\vert\right)\ ,
\label{eq:4a}
\ee
and spin matrices
\be
M_{\beta}(\hat{\bm e}) = (\sigma_0 +\beta {\bm\sigma}\cdot\hat{\bm e})\ ,
\label{eq:4b}
\ee
with $\hat e$ an arbitrary unit vector, we find
\be
G_k = \frac{1}{2} \sum_{\beta=\pm} F_k^{\beta}\,M_{\beta}\left(\frac{\vso{\bm k}-{\bm h}}{\vert \vso{\bm k}-{\bm h}\vert}\right)\ .
\label{eq:4c}
\ee
\ese
For a vanishing spin-orbit interaction the index $\beta$ turns into minus the spin-projection
index, and the spin-orbit interaction in zero field has an effect similar to that of a field in the absence of a
spin-orbit interaction. In particular, $\vso\neq 0$ splits the doubly degenerate band.
Now consider wave-vector convolutions of the Green function,

\bea
\varphi^{\beta_1 \beta_2}({\bm q},i\Omega_n) &=& \frac{1}{V} \sum_{\bm k} F_k^{\beta_1} F_{k-q}^{\beta_2} 
\nonumber\\
&& \hskip -70pt = \int \frac{d\Omega_{\bm k}}{4\pi}\,\frac{2\pi i \NF \sgn(\omega_m) \Theta\left(-\omega_m(\omega_m-\Omega_n)\right)}
                                                                    {i\Omega_n - \vF{\hat{\bm k}}\cdot{\bm q} + (\beta_2-\beta_1)\vert \vso\kF{\hat{\bm k}} - {\bm h}\vert} \hskip 20pt
\label{eq:5}
\eea
Here $\Omega_n$ is a bosonic Matsubara frequency, $q=(i\Omega_n,{\bm q})$, $d\Omega_{\bm k}$ is
the angular integration measure with respect to ${\bm k}$, and the second line represents the
leading contribution to the integral in the limit ${\bm q}, \Omega_n, {\bm h} \to 0$. These are the relevant ballistic
soft modes. We have derived them for noninteracting electrons, but interactions
cannot change their nature for reasons discussed in Refs.~\cite{Kirkpatrick_Belitz_2019a, Kirkpatrick_Belitz_2019b, Belitz_Kirkpatrick_2019}.
An inspection of Eq.~(\ref{eq:5}) reveals the following. Convolutions of quasiparticle resonances $F$
with different signs of the frequency (i.e., of retarded and advanced degrees of freedom) are soft as
${\bm q}, \Omega_n \to 0$ if $\beta_1 = \beta_2$. However, a magnetic field does not cut off this
singularity. These modes therefore cannot contribute to a nonanalytic dependence of the free
energy on the magnetic field or the magnetization. For $\beta_1 \neq \beta_2$, on the other hand,
the spin-orbit interaction gives the ballistic modes a mass even for ${\bm h} = 0$. For $\vso\neq 0$
there thus are no soft modes in a non-centrosymmetric system that can lead to a nonanalytic
free energy, and this is the source of the parameter $\nu$ in Eq.~(\ref{eq:1}) that cuts off the
nonanalyticity. This conclusion does not hinge on the particular form of the spin-orbit interaction
given in Eq.~(\ref{eq:3}); any spin-orbit interaction will split the band and give the soft modes a mass, so the
equation of state will have the same form.

This scenario for restoring a FM QCP in zero field is qualitatively different from the case of a gapless Dirac metal 
discussed in Ref.~\onlinecite{Belitz_Kirkpatrick_2019}, where the relevant soft modes exist, but do not couple to 
the order parameter. The class of candidate materials for this scenario is much smaller than for the one
discussed here, since it requires a special lattice symmetry. 
The current mechanism is also
very different from the effects of quenched disorder in the absence of a spin-orbit interaction: Disorder provides a 
mass under the logarithm in Eq.~(\ref{eq:1}) just as $\nu$ does, but it also leads to new soft modes that are
diffusive in nature and provide an additional nonanalytic contribution to the equation of state. 

In summary, we have shown that  non-centrosymmetric systems with a large spin-orbit coupling provide a 
platform for the realization of a FM QCP in clean systems in zero field, a goal that had eluded all 
experimental efforts for a long time. Two materials  in which this may already have been observed are
UIr and CeRh$_6$Ge$_4$, but more detailed studies of the quantum critical behavior are needed to
support this suggestion.

We conclude with a few comments about likely features of the resulting critical theory for a non-centrosymmetric
metal with a strong spin-orbit interaction. As mentioned after Eq.~(\ref{eqs:4}), the effects of
$\vso$ are similar to those of a magnetic field for $\vso=0$. As a result, the Gaussian vertex for the
3-component of the magnetization has two eigenvalues with the structure of Hertz theory for a ferromagnet,
with a dynamical exponent $z=3$, while the remaining eigenvalue has the structure of Hertz theory
for an antiferromagnet, with $z=2$, and the latter will lead to corrections to the
leading scaling behavior that results from the former. For instance, for the scaling of the critical
temperature with the control parameter $r = p - p_{\text{c}}$ for a pressure-tuned transition one
expects $T_{\text{c}} \propto (-r)^\tau$ with an effective exponent $\tau$ that is smaller than the
standard HMM value $\tau = 3/4$. This is consistent with a recent experiment that found
$\tau = 3/5$ in CeRh$_6$Ge$_4$ \cite{Kotegawa_et_al_2019}, but a more detailed investigation
is needed. More generally, it is not clear whether the QCP whose existence we have discussed, and
which Eq.~(\ref{eq:1}) provides a mean-field description for, is in the HMM universality class for some
systems, or for any systems. In particular, for Kondo lattice systems, such as CeRh$_6$Ge$_4$ \cite{Shen_et_al_2019, Kotegawa_et_al_2019},
questions arise about the interplay between the Kondo effect and quantum criticality. This topic has been
discussed predominantly for antiferromagnets \cite{Gegenwart_Si_Steglich_2008}, but various proposals
have been debated for ferromagnets as well \cite{Yamamoto_Si_2010, Komijani_Coleman_2018, Shen_et_al_2019}.
If HMM theory is not applicable, then the electrical resistivity in particular may not be governed by Mathon's $T^{5/3}$ law \cite{Mathon_1968},
but reflect a different ``strange metal'' behavior \cite{Shen_et_al_2019}. Also, FM analogs of the effects discussed for antiferromagnets in 
Ref.~\onlinecite{Woelfle_Abrahams_2011} might affect the critical behavior. These questions, as well as
the interplay of the spin-orbit interaction with quenched disorder, are open problems.

We thank Manuel Brando, Piers Coleman, Hisashi Kotegawa, and Uli Z{\"u}licke for discussions.


\vfill\eject

\onecolumngrid
\appendix
\centerline{\bf Supplemental Material for ``Ferromagnetic Quantum Critical Point in Non-Centrosymmetric Systems'':}
\bigskip
\centerline{\bf DERIVATION OF THE EQUATION OF STATE}
\bigskip
\bigskip
In this supplemental section we provide a derivation of the equation of state, Eq.~(1) in the main text. The logic of
the arguments is the same as in Ref.~\onlinecite{Brando_et_al_2016a_supp}.
\bigskip
\par
\centerline{\bf A: Magnetic order parameter coupled to conduction electrons}
\label{sec:I}
\medskip\par
Consider conduction electrons, described by Grassmannian spinor fields $\bar\psi$, $\psi$, in the presence of a fluctuating
magnetization described by a bosonic filed ${\bm m}$. The partition function for this coupled boson-fermion problem is
\be
Z = \int D[\bm m] D[\bar\psi,\psi]\,e^{-S[{\bm m},\bar\psi,\psi]}\ ,
\tag{S1}
\label{eq:S1}
\ee
and the action consists of three distinct parts,
\be
S[{\bm m},\bar\psi,\psi] = S_{\bm m}[{\bm m}] + S_{\text{F}}[\bar\psi,\psi] + S_{\text{c}}[{\bm m},\bar\psi,\psi]\ .
\tag{S2}
\label{eq:S2}
\ee
Here $S{\bm m}$ and $S_{\text{F}}$ describe the magnetization and the fermions, respectively, in isolation, and
$S_{\text{c}}$ describes the coupling between the two. The latter has the form of a Zeeman coupling,
\be
 S_{\text{c}}[{\bm m},\bar\psi,\psi] = -c \int dx\,{\bm m}(x)\cdot{\bm n}_{\text{s}}(x)
 \tag{S3}
 \label{eq:S3}
 \ee
 between the magnetization and the electronic spin density
 \be
 {\bm n}_{\text{s}}(x) = \bar\psi(x){\bm\sigma}\psi(x) \ .
 \tag{S4}
 \label{eq:S4}
 \ee
 Here $c$ is a coupling constant, 
 $x = ({\bm x},\tau)$ comprises the real-space position ${\bm x}$ and the imaginary-time variable $\tau$,
 and $\int dx = \int_V d{\bm x} \int_0^{1/T} d\tau$, with $V$ the system volume and $T$ the temperature. 
 The structure of the coupling term, Eq.~(\ref{eq:S3}), is completely general and reflects the fact that the spins of the conduction electrons will be
 subject to the effective field generated by the magnetization. It is valid independent of the origin of the
 magnetization, which can be due to the conduction electrons, or due to local moments, or a combination
 of the two \cite{Kirkpatrick_Belitz_2012b_supp, Brando_et_al_2016a_supp}.

 Now we treat the magnetization in a mean-field approximation, i.e., we replace the fluctuating field
 ${\bm m}$ by its average $\langle{\bm m}(x)\rangle = m\,{\hat z}$ that we take to point in the $z$-direction.
 The order-parameter action $S_{\bm m}$ then becomes a Landau theory, and the effect of the fermions
 can be described by formally integrating out the latter. This leads to a correction to the Landau action
 given by
 \be
 \delta S[m] = -\ln \int D[\bar\psi,\psi]\,e^{-S_{\text F}[\bar\psi,\psi] + cm \int dx n_{\text{s}}(x)} \propto \ln Z[m]
 \tag{S5}
 \label{eq:S5}
 \ee
 where $n_{\text{s}}$ is the $z$-component of ${\bm n}_{\text{s}}$, and 
 \be
 Z[m] = \langle e^{cm\int dx\,n_{\text{s}}(x)}\rangle_{\text F}
 \tag{S6}
 \ee
 and $\langle\ldots\rangle_{\text{F}}$ denotes an average with respect to the action $S_{\text{F}}$. 
 
 Now consider the spin susceptibility of the fermions in an effective magnetic field ${\mathfrak h}=cm$, which is given by
 the spin-density correlation function
 \begin{align}
 \chi(m) &= \frac{T}{V} \int dx\,dy\, \left\langle\left(n_{\text{s}}(x)-\langle n_{\text{s}}(x)\rangle_{S_{\mathfrak h}}\right)\left(n_{\text{s}}(y)-  \langle n_{\text{s}}(y)\rangle_{S_{\mathfrak h}}\right)
                       \right\rangle_{S_{\mathfrak h}}
 \nonumber\\
 &=  \frac{T}{V} \int dx\,dy\, \left\langle n_{\text{s}}(x)\,n_{\text{s}}(y)\right\rangle_{S_{\mathfrak h}} - \frac{V}{T} \left(\langle n_{\text{s}}(x)\rangle_{S_{\mathfrak h}}\right)^2
\tag{S7}
 \label{eq:S7}
 \end{align}
 of fermions governed by an action
 \be
 S_{\mathfrak h} = S_{\text{F}} + {\mathfrak h} \int dx\,n_{\text{s}}(x)
 \tag{S8}
 \label{eq:S8}
 \ee
 Now we observe
 \be
 d\ln Z[m]/dm = \frac{V}{T}\,c\, \langle n_{\text{s}}(x)\rangle_{S_{\mathfrak h}}
 \tag{S9}
 \label{eq:S9}
 \ee
 and 
 \be
 d^2\ln Z[m]/dm^2 = \frac{V}{T}\,c^2\,\chi(m)
 \tag{S10}
 \label{eq:S10}
 \ee
 as well as $\ln Z[m=0]=\ln 1 = 0$ and $(d\ln Z[m]/dm)_{m=0} = c\int dx \langle n_{\text{s}}(x)\rangle_{\text F} = 0$. Equations~(\ref{eq:S10}) and (\ref{eq:S5})
 therefore imply
 \be
 \delta S[m] \propto \frac{-V}{T}\,c^2\int_0^m dm_1 \int_0^{m_1} dm_2\,\chi(m_2)
 \tag{S11}
 \label{eq:S11}
 \ee
 The renormalized mean-field theory for the magnetization in the presence of a physical magnetic field $h$ that incorporates the coupling of the conduction 
 electrons to the magnetic order parameter is thus given by an effective action
 \be
 S_{\text{eff}}[m] = \frac{r}{2}\,m^2 + \frac{u}{4}\,m^4 + O(m^6) - h m + \delta S[m]
 \tag{S12}
 \label{eq:S12}
 \ee
 
 \bigskip
\par
\centerline{\bf B: The spin susceptibility, and the renormalized mean-field equation of state}
\label{sec:II}
\medskip\par

 In Sec.~A we saw that we need the magnetic-field dependence spin susceptibility of the conduction electrons as input.
 In an ordinary Landau Fermi liquid the spin susceptibility is a nonanalytic function of the field, see Refs.~\onlinecite{Belitz_Kirkpatrick_Vojta_1997_supp, 
 Betouras_Efremov_Chubukov_2005_supp, Brando_et_al_2016a_supp} 
 and the discussion after Eq.~(5) in the main text. In three spatial dimensions, and at zero temperature, the leading nonanalyticity is
 \be
 \chi(h\to 0) = \chi(h=0) - \chi_2\,h^2\ln(h)
 \tag{S13}
 \label{eq:S13}
 \ee
 with $\chi_2>0$ a positive coefficient. This nonanalyticity is the result of soft two-particle excitations in the Fermi liquid that are rendered massive
 by a magnetic field and also by a nonzero temperature.  A spin-orbit interaction also gives the soft modes a mass, as explained in the context
 of Eq.~(5) in the main text. In a centrosymmetric Dirac Fermi liquid, as defined in the main text and in 
 Refs.~\onlinecite{Kirkpatrick_Belitz_2019a_supp, Kirkpatrick_Belitz_2019b_supp, Belitz_Kirkpatrick_2019_supp}, chiral degrees of freedom provide 
 a new class of soft modes, and the spin susceptibility is still given by Eq.~(\ref{eq:S13}) \cite{Kirkpatrick_Belitz_2019a_supp}.
 However, in the absence of spatial inversion symmetry no such compensation mechanism is available. 
 With $h$, $t$, and $\nu$ the magnetic field, temperature, and spin-orbit coupling strength in atomic units, we then have
 \be
 \chi(m) = \text{const.} - (\chi_2/2) \ln(m^2 + t^2 + \nu^2)
 \tag{S14}
 \label{eq:S14}
 \ee
 Equations~(\ref{eq:S11}, \ref{eq:S12}), and (\ref{eq:S14}) then yield an equation of state that has the structure of Eq.~(1) in the main text.

\end{document}